\documentclass[showpacs,prb,preprint]{revtex4}
\usepackage{graphicx}

\begin{document}

\title{ Bistability in a magnetic and nonmagnetic double-quantum-well
structure mediated by the magnetic phase transition}
\author{ Y. G. Semenov, H. Enaya, and K. W. Kim}
\address{Department of Electrical and Computer Engineering, North
Carolina State University, Raleigh, NC 27695-7911}

\begin{abstract}
The hole distribution in a double quantum well (QW) structure
consisting of a magnetic and a nonmagnetic semiconductor QW is
investigated as a function of temperature, the energy shift
between the QWs, and other relevant parameters. When the itinerant
holes mediate the ferromagnetic ordering, it is shown that a
bistable state can be formed through hole redistribution,
resulting in a significant change in the properties of the
constituting magnetic QW (i.e., the paramagnetic-ferromagnetic
transition). The model calculation also indicates a large window
in the system parameter space where the bistability is possible.
Hence, this structure could form the basis of a stable memory
element that may be scaled down to a few hole regime.
\end{abstract}

\pacs{72.20.Ht,85.60.Dw,42.65.Pc,78.66.-w}
\maketitle


Spintronics~\cite{spinel} approaches the brink of applications in the micro-
and nano-electronics. As soon as the semiconductor compounds doped with
magnetic ions reveal room temperature ferromagnetism, the creation of
competitive electronic devices will become the barest necessity. The study
of different structures and their spin effects, which could result in spin
devices at lower temperatures, is already a topical problem especially when
we take into the account the continued improvement in the critical
temperature $T_{c}$ of the paramagnetic-ferromagnetic (PM-FM) phase
transition.

In this Letter, we explore the formation of a bistable state and
its device application in a double quantum well (QW) semiconductor
structure with respect to the distribution of itinerant holes
between the constituting nonmagnetic and magnetic QWs. The
calculation illustrates that the interplay between the free
carriers and the magnetic ion spins is the key to achieving the
bistability where the magnetic QW can switch between the PM and FM
phases in a controlled manner.

The structure under investigation is illustrated in Fig.~1.
Consider a magnetic QW (MQW) and a nonmagnetic QW (NQW) of widths
$L_{wM}$ and $L_{wN}$, respectively, separated by a barrier. The
width $L_{b}$ and the hight of the barrier should be large enough
to form non-coherent single QW states and yet not too large in
order to enable hole redistribution when a gate bias is applied.
The total 2D hole concentration $n_{h}^{0}$ is assumed to be a
constant and the FM-PM transition in the MQW to be mediated by
two-dimensional (2D) free holes. If the hole energy in the NQW is
lower than that in the MQW, one can expect that the stable state
will correspond to hole localization primarily in the NQW with a
small leakage in the MQW, which is in a PM phase (see the
schematic on the left in Fig.~1). When a proper bias is applied,
holes from the NQW can be transferred to the MQW via tunneling,
over-barrier injection, etc. As the hole density in the MQW
surpasses a certain threshold at a given operating temperature,
the layer undergoes the PM-FM transition. When the hole exchange
interaction with the FM-ordered ion spins in the MQW is strong, it
can reduce the total free energy below that of the initial state
with the PM phase even after the bias is switched off. Then, the
holes will remain confined in the MQW and the FM state be
maintained (the schematic on the right in Fig.~1). When a reverse
bias pulse is applied, the holes are drained out of the MQW into
the NQW, and the MQW will return to the PM state. Hence, if
realized, these two stable states can coexist under the same
external condition and the switching between them is mediated by
the electrically-controlled PM-FM phase transition.~\cite{Ohno} It
is expected that the structure can operate up to a temperature
slightly below the saturated maximum of $T_{c}$, which can reach
room temperature or higher in some material systems.~\cite{Tsui}

To analyze this problem, typically one would derive the magnetic
Hamiltonian $H_{m}$, which describes the effect of the effective
ferromagnetic inter-ion spin-spin interaction in the presence of
free holes. Then the calculation of the free energy $F$ of the
total system consisting of magnetic ions and free carriers, which
occupy the MQW with a concentration of $n_{hM}$ and the NQW with
$n_{hN}$ ($n_{hM}+n_{hN}=n_{h}^{0}$), leads to the carrier
population factor $\eta =n_{hM}/n_{h}^{0}$ and $1-\eta
=n_{hN}/n_{h}^{0}$ at each QW that minimizes $F=F(\eta )$. The
existence of two local minima with respect to  $\eta $
demonstrates the bistability of the system under consideration.

Although conceptually correct, this approach faces the difficulty
of specifying the mechanisms responsible for FM ordering in the
MQW (for details on various mechanisms, refer to those cited in
Ref.~\onlinecite{PashRyachetal}). To circumvent this problem,
which is beyond the scope of the present study, we develop a
semi-phenomenological approach that utilizes the data extracted
from routine experimental measurements of magnetism. Namely, we
assume that the magnetic part of the free energy $ F_{M} $ can be
expanded with the magnetization $M$ (order parameter) according to
Landau theory. This expansion approximates satisfactorily the
veritable dependence in the whole temperature interval we are
interested in. Assuming the easy magnetization axis is directed
along the growth axis of the sample and a magnetic field
$\overrightarrow{B}$ parallel to $ \overrightarrow{M}$, the
$F_{M}$ expansion in the most general form reads
\begin{equation}
F_{M}=-a(T_{c}-T)M^{2}+bM^{4}-\frac{1}{2}MB.  \label{f1}
\end{equation}%
It can be shown that the parameters $a$ and $b$ of Landau
expansion in Eq.~(\ref{f1}) can be expressed in terms of
fundamental properties of the magnet: The Curie-Weiss law for the
magnetic susceptibility $\chi =C_{0}/(T-T_{c})$ at $T>T_{c}$
defines $\ a=1/4C_{0}$, while the spontaneous magnetization $
M_{s} = [a(T_{c}-T)/2b]^{1/2} = M_{0}\sqrt{1-T/T_{c}}$ for
$T<T_{c}$ (that minimizes $F_{M}$ at $B=0$) provides
$b=aT_{c}/2M_{0}^2$. By these relations, all parameters in $F_{M}$
in Eq.~(\ref{f1}) are determined. $ T_{c} $ can be found
separately as a function of the magnetic layer carrier population.

Since we are looking for the free hole distribution over the
magnetic and nonmagnetic QWs at $B=0$, the magnetization $M$ in
Eq.~(\ref{f1}) takes the equilibrium value. Hence,
\begin{equation}
F_{M}=-\frac{a^{2}(T_{c}-T)^{2}}{4b}, ~ T<T_{c};~~~~ F_M=0, ~ T>T_{c}.
\label{f2}
\end{equation}
It is important to note that Eq.~(\ref{f2}) includes the
dependence on the free hole concentration in the MQW via the
critical temperature $ T_{c}=T_{c}(\eta )$.

The total free energy of the system can be obtained if Eq.~(\ref{f2}) is
supplemented with the free energy of the hole gas:
\begin{equation}
F_{h}=F_{2D}(\eta )+U(\eta )+C(\eta ).  \label{f3}
\end{equation}
The first term in Eq.~(\ref{f3}) accounts for the kinetic energy
of the hole gas in both QWs, $U(\eta )$ is the hole potential that
is different for the MQW and the NQW, and $C(\eta )$ is the energy
of the Coulomb interaction between the QWs. Assuming the parabolic
dispersion law with an effective mass $m$ for 2D holes and that
only the lowest sub-bands of the QWs can be populated by holes
with $L_{wM} = L_{wN}$, one can find the kinetic energy for the
hole gas as
\begin{equation}
F_{2D}(\eta)=k_{B}T\left\{\eta f_{1}\left(\frac{\varepsilon_{F}^{0}}{k_{B}T}%
\eta\right) +(1-\eta)f_{1}\left( \frac{\varepsilon
_{F}^{0}}{k_{B}T} (1-\eta )\right)\right\} n_{h}^{0},  \label{f4}
\end{equation}
where $k_{B}$ is the Boltzmann constant, $\varepsilon _{F}^{0}=\pi
\hbar ^{2}n_{h}^{0}/m$ is the Fermi energy of 2D holes with
concentration $ n_{h}^{0}$, and
\begin{equation}
f_{1}(x)=\ln \left( e^{x}-1\right) +\frac{1}{x}Li_{2}\left( 1-e^{x}\right)
\label{f4a}
\end{equation}
with a polylogarithmic function $Li_{2}\left( x\right)
=\int_{x}^{0} dt\ln (1-t)/t$. At low temperature $T\ll \varepsilon
_{F}^{0}/k$, Eq.~(\ref{f4}) describes the sum of degenerate
carrier energies in both QWs. The low temperature assumption is
commonly used in the works on the FM ordering in MQWs. For the
present purpose, however, we need to account for the arbitrary
relation between $\varepsilon _{F}^{0}$ and $k_{B}T$ according to
Eq.~(\ref{f4}). The contribution of the energy shift $\Delta U$
between the MQW and
the NQW (Fig.~1) is accounted for in the term%
\begin{equation}
U(\eta )=\Delta U\eta n_{h}^{0},  \label{f5}
\end{equation}
while the Coulomb energy in the strong confinement limit takes the form
\begin{equation}
C(\eta )=\frac{2\pi \text{\textbf{e}}^{2}}{\epsilon }L_{b} {
n_{h}^{0}}^2 \left( \eta -\frac{1}{2}\right) ^{2},  \label{f6}
\end{equation}
where \textbf{e} is the electron charge and $\epsilon $ the dielectric
constant.

Now one can analyze the total free energy $F=F_{M}+F_{h}$ with
respect to possible FM phase transitions in the MQW. Considering
that the realistic dependence $T_{c}=T_{c}(n_{hM})$ should be
taken from the experiments, we specify our analysis by a model
that can be applied to a typical diluted magnetic semiconductor.
Assuming that the total magnetization of the MQW stems mainly from
the magnetic ions and $n_{h}^{0}\ll n_{m}L_{wM}$ ($n_{m}$ is the
3D magnetic ion concentration in the MQW), one can easily find the
parameters $a=3/[4S(S+1)g^{2}\mu _{B}^{2}n_{m}]$ and $ b=
3k_{B}T_{c}/[8S^{3}(S+1)g^{4}\mu _{B}^{4}n_{m}^{3}]$, where $g$ is
the magnetic ion $g$-factor with spin $S$ and $\mu _{B}$ denotes
the Bohr magneton.

In order to describe the dependence of $T_{c}$ on $\eta $, we propose an
approximation
\begin{equation}
T_{c}=T_{c}^{0}\left( 1-e^{-\alpha \varepsilon _{F}^{0}\eta
/k_{B}T_{c}^{0}}\right) ,  \label{f7}
\end{equation}%
where $T_{c}^{0}$ is the asymptotic (at a high enough $n_{hM}$)
value of the critical temperature and $\alpha $ is the fitting
parameter that adjust the dependence [Eq.~(\ref{f7})] to the
experiments. In the following, we assume $ \alpha =1$ since it
describes the experimental results satisfactorily.~\cite{Boukari}
Combining this approximation with Eqs.~(\ref{f2}), (\ref{f4}),
(\ref{f5}), and (\ref{f6}), we find the trial function in the form
of the free energy per hole normalized by the energy unit
$k_{B}T_{c}^{0}$
\begin{eqnarray}
F &=&-\frac{3}{8}\frac{S}{S+1}\frac{\nu }{t_{c}(\eta )}\left[ t_{c}(\eta )-t%
\right] ^{2}\theta (t_{c}(\eta )-t)  \nonumber \\
&&+t\left\{ \eta f_{1}\left( \frac{r}{t}\eta \right) +\left( 1-\eta \right)
f_{1}\left( \frac{r}{t}\left( 1-\eta \right) \right) \right\}   \nonumber \\
&&+u\eta +w(2\eta -1)^{2}+F_{ex},  \label{f8}
\end{eqnarray}%
where $\nu =n_{m}L_{w}/n_{h}^{0}$, $t=T/T_{c}^{0}$, $t_{c}(\eta
)=T_{c}/T_{c}^{0}$, $u=\Delta U/k_{B}T_{c}^{0}$, $w=\pi $\textbf{e}$%
^{2}L_{b}n_{h}^{0}/2\epsilon k_{B}T_{c}^{0}$, $r=\varepsilon
_{F}^{0}/k_{B}T_{c}^{0}$, $t_{c}(\eta )=1-\exp (-r\eta )$, and $\theta (x)$
is the Heaviside step function. For completeness, Eq.~(\ref{f8}) also
includes the exchange energy of free carriers $F_{ex}$, which can influence
the effect of bistability.~\cite{WuAPL,Sarma97} We performed the calculation
of $F_{ex}$ in a Hartree-Fock approximation following Ref.~%
\onlinecite{Phatisena}, where the exchange and the correlation potentials of
a 2D gas were found at finite temperatures. The final result of our
calculation has a proper analytical approximation in terms of the
dimensionless parameter $r$
\begin{eqnarray}
F_{ex} &\cong &-\frac{2^{5/2}\mathbf{e}^{2}\sqrt{n_{h}^{0}}}{3\sqrt{\pi }%
\epsilon k_{B}T_{c}^{0}}\times   \nonumber \\
&&\left\{ \eta ^{3/2}\phi \left( \frac{r}{t}\eta \right) +(1-\eta
)^{3/2}\phi \left( \frac{r}{t}(1-\eta )\right) \right\} ,  \label{f9}
\end{eqnarray}%
where $\phi \left( x\right) =1-e^{-x/2.4}+x^{b}/(2.4+x^{c})$, $%
b=0.3+0.2\theta (x-1)$, $c=2.2-0.2\theta (x-1)$. Note that Eq.~(\ref{f8})
takes a form similar to the free energy expression used for analyzing the
spin/charge separation of magnetic semiconductors near a PM-FM phase
transition.~\cite{Guinea} 

For a numerical evaluation, let us assume the following "typical"
values for the parameters of the double QW structure: $m=0.3m_{0}$
($m_{0}$ is the free electron mass), $\epsilon =12.9$, $S=5/2$,
$L_{wM}=L_{wN}=10$~nm, $L_{b}=5$~nm, $n_{h}^{0}=10^{12}~ {\rm
{{cm}^{-2}}}$, $n_{m}=1.3\times 10^{21}~ {\rm {{cm}^{-3}}}$, and
$T_{c}^{0}=100$~K. Figure~2 displays $F(\eta )$ calculated at
three different values of the energy shift $\Delta U$ between the
minima of PM QW and NQW ($u=\Delta U/k_{B}T_{c}^{0}=29,5,-9$).
Clearly, curves 1 and 3 ($u=29,-9$) support only one stable state
at $\eta =0$ or 1. This means that when the MQW state lies either
too high (curve 1) or too low (curve 3) compared to that of the
NQW, the holes strongly prefer to be confined in one of the QWs.
Even when they are transferred to the other QW through an external
bias, the holes will return to the preferred state once the bias
is turned off. However, one can realize a structure that has free
energy minima at or near both $\eta =0$ (with the MQW in the PM
phase) and $\eta =1$ (the MQW in the FM phase) if $\Delta U$ is
properly selected (curve 2). These two states can be stable with
respect to small fluctuations under the same external conditions.
We also found that the relative effect of $F_{ex}$ is small
compared to the Coulomb energy of the inter-QW carrier interaction
$w(2\eta -1)^{2}$.

This bistability can be achieved in a relatively wide range of
$\Delta U$ and $T$ as shown in Fig.~3. As expected, the condition
for $\Delta U$ becomes less stringent with a decreasing $T$. This
is because the structure can now operate with a lower $T_C$, which
in turn requires a smaller hole density in the MQW for the PM-FM
transition. The highest operating temperature will be somewhat
lower than $T_{c}^{0}$ and is a function of the maximum possible
$n_{hM}$.

It should be emphasized again that the structure is analyzed with
respect to the free hole distribution $\eta $ that minimizes the
system free energy. Hence, the height of the local maximum in the
free energy (e.g., near $\eta =0.8$ for curve 2 in Fig.~2) does
not constitute the energy barrier separating the NQW and the MQW.
In fact, details of the barrier layer shown in Fig.~1 is not
considered in the present study. Rather, the stability of local
minima (thus, the lifetime) depends explicitly on the magnitude of
the fluctuation in $\eta $ that can cause unwanted switching.
Since the required value is large according to our calculation
(e.g., $ \Delta \eta \gtrsim 0.2) $, the system is expected to be
robust against most fluctuations including thermal transitions.
Hence, the lifetime at each local minima may be as long as that of
conventional magnetic memory cells.  A detailed analysis is
necessary for a quantitative estimate.

In summary, we demonstrate bistability formation in the structure
consisting of magnetic and nonmagnetic QWs. The bistability is
mediated by the PM-FM transition in the MQW. In contrast to the
case of coupled NQWs,~\cite{WuAPL,Sarma97} our structure reveals a
bistability as soon as the conditions for the ferromagnetism in
the MQW are satisfied. As a result, a room temperature operation
may be achieved once a proper magnetic material is developed.
Although the investigation is done for a 2D structure, it is
expected that a similar principle can also be applied to the 0D
system.  As the size of the gate electrode shrinks, the MQW can
form a gated quantum dot, which can undergo the PM-FM (or
super-PM) transition by controlling population/depopulation of
holes.  Furthermore, one can envision a magnetic nanocrystal
embedded in a nonmagnetic barrier in place of the MQW.  Hence,
this structure can form the basis of a stable memory element that
may be scaled down to a few hole regime with very low power
consumption.

This work was supported in part by the DARPA/ONR and the SRC/MARCO Center on
FENA.

\newpage \noindent Figure captions

\vspace{0.2 in} \noindent Fig. 1. Schematic energy diagram (valence band) of
the structure in the two coexisting stable states. The mutual alignment of
the hole spins (large arrows) and localized spins (small arrows) reduces the
energy in the FM QW by $E_{exch}\simeq2F_{M}/n_{h}^{0}$ due to their
exchange interaction. The switching between these states (arch arrows) is
achieved by applying an appropriate bias pulse on the gate electrode.

\vspace{0.2 in} \noindent Fig. 2. Free energy trial function $F(\eta)$ at
different values of $u$ ($=\Delta U/k_{B}T_{c}^{0}$). Three different
scenarios are shown: curve 1 ($u=29$) - a monostable case with holes
occupying the NQW ($\eta=0$), while the MQW is in a PM phase; curve 2 ($u=5$%
) - a bistable case where the PM and FM phases coexist; curve 3 ($u=-9$) - a
monostable case with holes populating the MQW ($\eta=1$) in the FM phase.

\vspace{0.2in} \noindent Fig. 3. Phase diagram of the parameter
space indicating the potential bistability region.  A free hole
density $n_{h}^{0} $ of $ 10^{12} ~{\rm {cm^{-2}}}$ is assumed.

\newpage

\begin{center}
\begin{figure}[tbp]
\includegraphics[scale=.4,angle=-90]{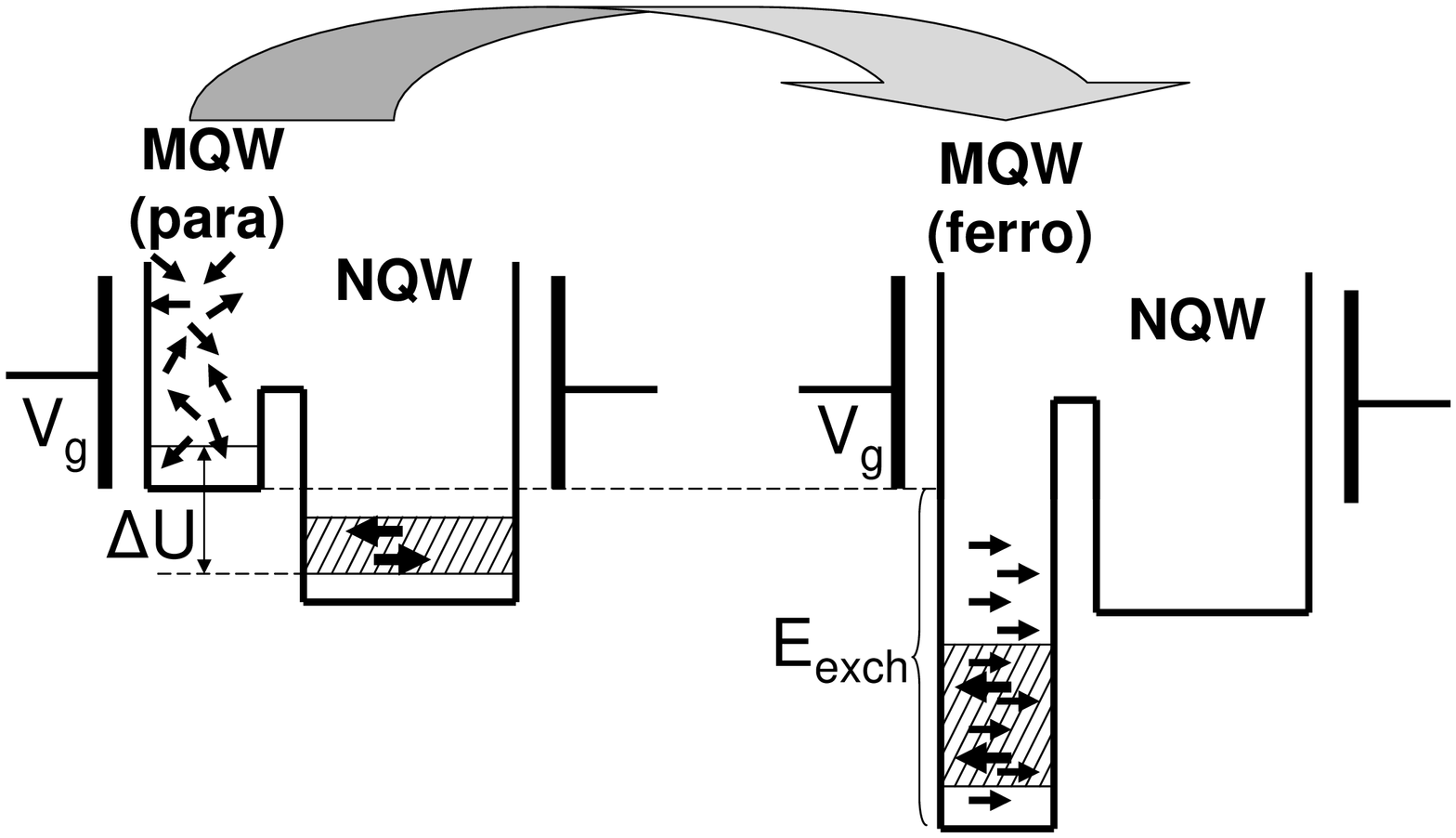}
\end{figure}
\vspace{320pt} {\large Fig. 1: Semenov et al. }
\end{center}

\newpage

\begin{center}
\begin{figure}[tbp]
\includegraphics[scale=.7]{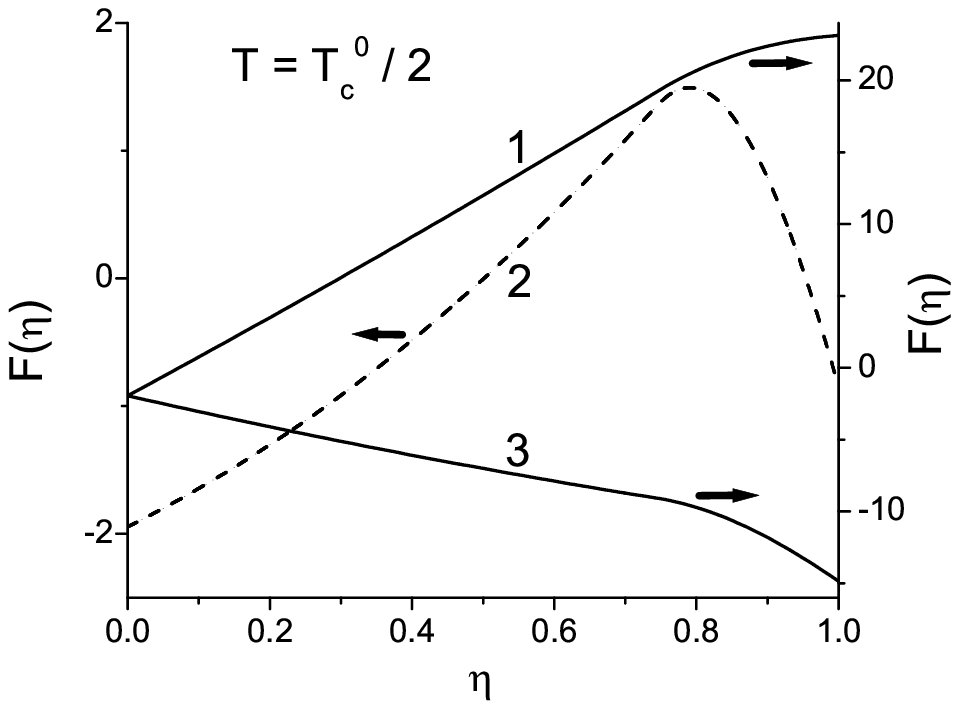}
\end{figure}
\vspace{320pt} {\large Fig. 2: Semenov et al. }
\end{center}

\newpage

\begin{center}
\begin{figure}[tbp]
\includegraphics[scale=.7]{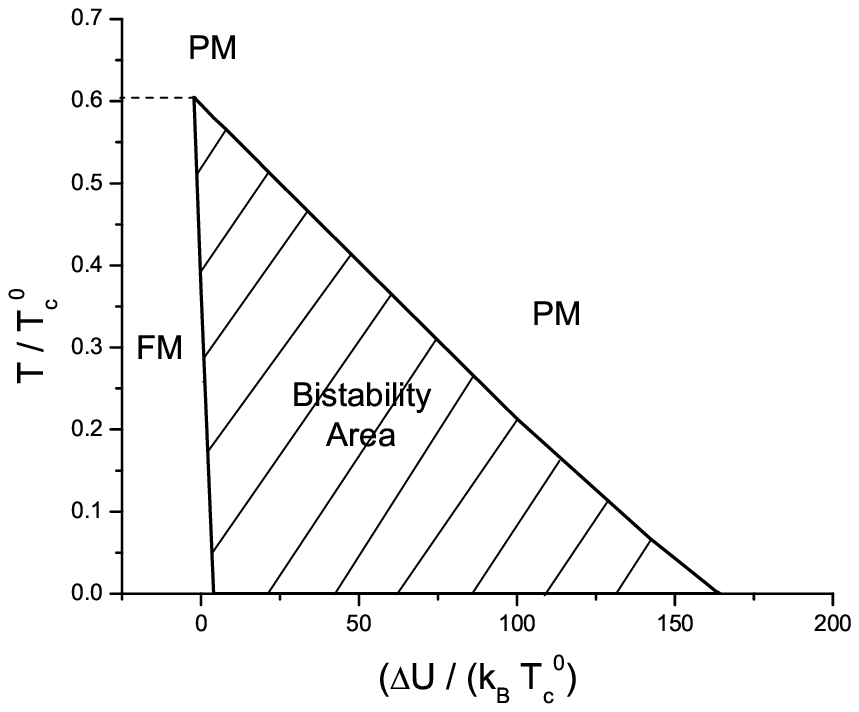}
\end{figure}
\vspace{320pt} {\large Fig. 3: Semenov et al. }
\end{center}

\end{document}